\documentclass[10pt]{article}
\usepackage[a4paper, total={6in, 8in}]{geometry}
\usepackage{color}
\usepackage{hyperref}
\usepackage{csquotes}

\usepackage{amssymb}

\usepackage[figuresright]{rotating}

\begin{document}





\title{Quantitative study about the estimated impact of the AI Act}











\author{Marc P. Hauer\\ Tobias D. Krafft\\ Dr. Andreas Sesing-Wagenpfeil\\ Prof. Katharina Zweig}

\maketitle

\begin{abstract}
With the Proposal for a Regulation laying down harmonised rules on Artificial Intelligence (AI Act) the European Union provides the first regulatory document that applies to the entire complex of AI systems. While some fear that the regulation leaves too much room for interpretation and thus bring little benefit to society, others expect that the regulation is too restrictive and, thus, blocks progress and innovation, as well as hinders the economic success of companies within the EU.
Without a systematic approach, it is difficult to assess how it will actually impact the AI landscape.

In this paper, we suggest a systematic approach that we applied on the initial draft of the AI Act that has been released in April 2021. We went through several iterations of compiling the list of AI products and projects in and from Germany, which the Lernende Systeme platform lists, and then classified them according to the AI Act together with experts from the fields of computer science and law.
Our study shows a need for more concrete formulation, since for some provisions it is often unclear whether they are applicable in a specific case or not. Apart from that, it turns out that only about 30\% of the AI systems considered would be regulated by the AI Act, the rest would be classified as low-risk.
However, as the database is not representative, the results only provide a first assessment.

The process presented can be applied to any collections, and also repeated when regulations are about to change. This allows fears of over- or under-regulation to be investigated before the regulations comes into effect.

\end{abstract}

\section{Introduction}

In February 2020, two of the authors were participants of the Enquete commission meeting of the German parliament. Their talk about the need for regulation of AI systems, was greeted by strong adversarial arguments that any kind of regulation would be likely to overregulate the area of AI development and application. However, numbers were missing on both sides: how many AI systems in Germany would be subject to which kind of regulation? 
While the 'Proposal for a Regulation laying down harmonised rules on Artificial Intelligence'\footnote{Version Brussels, 21.4.2021 COM(2021) 206 final 2021/0106(COD)} -- hereafter referred to as 'AI Act' -- is still not in its final version, this paper aims to present a method and some first, quantitative numbers on the percentage of AI systems which are covered by at least one of the risk classes which are explicitly set out in the AI Act.

The development of the method also revealed some cases which do not seem to be addressed by the AI Act - the method might thus be helpful to identify border cases or help to re(de)fine the current classification. 
Finally, our method showed that it is indispensible  to work in interdisciplinary teams to categorize AI systems into the risk-categories: Some cases could only be identified as high risk by the involved computer scientists while others had to be identified by the involved legal scientist. 

It is to note that, due to several amendments to the initial draft of the AI Act from April 2021 till December 2022, our study has been carried out with respect to the initial draft. Therefore, we could roughly consider the last consolidated version\footnote{Version Brussels, 25.11.2022 14954/22 2021/0106(COD)} of the AI Act that introduces an additional class that was not yet available for our study. Use cases that belong to that new class could not be thoroughly assessed before, thus they have been excluded from our study during the preprocessing phase. 


\section{The risk classes of the current AI act proposal}
The AI Act, like numerous other proposals for regulating AI systems, such as the Ethics Guidelines of the High Level Expert Group on Artificial Intelligence~\cite[p.4]{ai2019high}, follows a risk-based approach for the classification of AI systems. This approach is the result of considering several policy options for taking regulatory action on AI. The European Commission has, according to its own statements, based its decision, among other things, on the economic and social impact of the planned regulation, with a particular focus on the impact on fundamental rights. Thus, the proposed risk-based approach also aims to ensure the proportionality of the proposed regulation.\footnote{See Recital No. 14 AI Act.} The result of these considerations is a classification of AI systems into a total of five risk classes. In the initial draft of the AI Act as of April 2021, three of these classes are explicitly mentioned, whilst the fourth is an implicit one: the explicit risk classes are prohibited AI systems (1.), high-risk AI systems (2.) and \enquote{other AI systems}, which can be described as AI systems with special needs for transparency (3.). The fourth risk class referred to is the class of AI systems that, under the AI Act's approach, are not subject to legal regulation and can therefore be described as low-risk AI systems (4.). It should be noted that the classification as an AI system with special needs for transparency does not exclude the classification as a prohibited AI systems or high-risk AI system at the same time. As a result of the intense debate on the AI Act during the ongoing legislative procedure, an additional fifth class of \enquote{general purpose} AI systems has been added in the latest proposal (5.).

\subsection{Prohibited AI systems}
The most explicit types of systems are the forbidden AI systems. They are described in Art 5 of the AI Act. Art. 5(1) of the draft contains a list of prohibited AI practices. This concerns a total of four use cases for AI systems.

Art. 5(1)(a) of the AI Act covers AI systems that use subliminal techniques to influence the behavior of a person in a way that causes or may cause physical or psychological harm to that person or to another person. This refers to techniques that influence people unconsciously or subconsciously. These systems may not be placed on the market, put into operation or used within the EU.

According to Art. 5(1)(b), AI systems that exploit a weakness of a certain group of persons due to their age or a physical or mental disability in order to significantly influence the behavior of a person belonging to this group in a way that causes or may cause physical or mental harm to this person or to another person shall be prohibited. The placing on the market, putting into service and use of such devices is also prohibited.

Prohibited under Art. 5(1)(c) is the placing on the market, putting into service and use of AI systems for the purpose of evaluating or classifying natural persons over a certain period of time based on their social behavior or known or predicted personal or personality characteristics (social scoring). In order for a system to fall under the prohibition, it is additionally required that the social assessment either leads to the worse treatment or disadvantage of certain natural persons or groups of persons in social contexts that have no connection with the circumstances under which the data was originally collected or that the explained worse treatment is unjustified or disproportionate with regard to the social behavior of the data subjects or its gravity.

Article 5(1)(d) prohibits in principle the use of real-time biometric remote identification systems in publicly accessible areas for law enforcement purposes, with three exceptions: targeted search for potential victims of crime; prevention of danger to life or limb or to critical infrastructure; tracing of perpetrators who are wanted on a European arrest warrant for the commission of serious crimes. In contrast to the other systems described in Art. 5(1), only the use, but not the placing on the market or putting into service, is prohibited. This should probably take into account the fact that the purpose disapproved by Art. 5(1)(d) is not intrinsic to the system, but results from the specific purpose of the use of real-time remote identification systems, which is determined exclusively by the operator of the system.

\subsection{High-risk AI systems}
The second and - measured by the scope of the AI Act - most significant type of systems, are the so-called high-risk AI systems. This term is not defined as such, but the classification results from the determination of the requirements for the existence of an AI system in Art. 6 AI Act. Two alternatives are to be distinguished here.

According to Art. 6(1) of the initial draft AI Act, an AI system may be considered a high-risk AI system under two conditions. According to letter a), the AI system must either be used as a safety component of a product that is subject to one of the sector-specific product safety rules harmonized in Annex II to the AI Act, or the AI system must itself constitute a product within the scope of application of the legislation listed in Annex II. An AI system that meets this requirement shall be considered a high-risk AI system under point (b) only if the product of which the AI system is a safety component or the AI system itself is subject to third-party conformity assessment under the provisions of the acts referenced by Annex II. The latest published version contains, in principle, the same distinction: Art. 6(1) names AI systems which are themselves a product within the scope of the listed legislation in Annex II, whereas Art. 6(2) names AI systems that are intended to be used as a safety component of a product covered by the legislation referred to in Annex II. As we have conducted our study on the basis of the initial draft, we have not taken into account the new order of the alternatives set out in Art. 6 AI Act.

In addition and independently of the requirements of Art. 6 para. 1 and para. 2, paragraph 3 of this Article stipulates that AI systems listed in Annex III to the AI Act shall be considered high-risk.\footnote{The latest draft of the AI Act contains an exemption for those cases in which the output of a system is purely accessory in respect of the relevant action or decision to be taken; as mentioned above, we could not consider this exemption during our study.} Annex III contains an exhaustive list of eight areas of application for AI systems: (1) biometrics; (2) critical infrastructure; (3) education and vocational training; (4) employment, workers management and access to self-employment; (5) access to and enjoyment of essential private services and essential public services and benefits; (6) law enforcement; (7) migration, asylum and border control management; (8) administration of justice and democratic processes.

However, the mere use of an AI system within one of these application areas does not lead to its classification as a high-risk AI system. Rather, Annex III identifies in detail use cases for AI systems that are to be considered high-risk AI systems and assigns them to the eight application areas.\footnote{Due to some minor amendments to the list of use cases in Annex III, our study does not cover the full scope of use cases set out in the last version of the AI Act. This concerns the addition of "critical digital infrastructure" as well as AI systems intended to be used for risk assessment and pricing in the case of life and health insurance.}

The following are examples of use cases for AI systems that are explicitly classified as high-risk AI systems\footnote{In this paper, specific entries are referred to with their ID. Appendix~\ref{ap:use_cases} shows the respective full texts.}:
\begin{itemize}
    \item Annex III 2a: Automatic online classification of critical power grid situations (ID 273)
    \item Annex III 3a: Intelligent tutor for technology-enhanced learning (ID 655)
    \item Annex III 4a: Anonymous Predictive People Analytics (ID 414)
\end{itemize}

Other use cases will be discussed in more detail later when open and unclear aspects of the classification are discussed. It is to note that, in particular, the rationale behind the classification as a high-risk AI system is not part of the definition. It can only be derived indirectly from Art. 7 AI Act. According to the initial proposal of the AI Act, this provision authorizes the EU Commission to supplement the list of Annex III and, in this context, determines the conditions under which use cases of AI systems not previously covered may be included in Annex III. Accordingly, the addition is permissible under two conditions. First, the newly included AI system must belong to one of the fields of application already listed in Annex III (cf. Art. 7(1)(a) AI Act). The authority of the EU Commission is thus limited: The Commission is not authorized to supplement the existing catalog with additional areas of application. 

The second requirement determines that an AI system may only be included by the Commission in the catalog of high-risk systems under Annex III if it poses a risk of harm to health or impairment of safety or a risk of adverse effects on fundamental rights and, in addition, this risk is equivalent or greater than the risks of those systems already listed in Annex III in terms of its severity and probability of occurrence (cf. Art. 7(1)(b) AI Act). The comparison must be made with the individual use cases in accordance with the purpose of the provision, as AI systems with different risk profiles are conceivable in the enumerated areas of application. Art. 7(2) AI Act then specifies the factors to be taken into account when determining the risks posed by an AI system; these are, among others, the purpose of the AI system, the extent of the (anticipated) use of the AI system, the extent to which damage has already occurred, and the intensity of the damage that has already occurred or is expected to occur.

Already here, some weaknesses of the previous draft of the AI Act can be identified, of which further will be discussed later in detail (see Section~\ref{sec:discussion}). For example, the text of Art. 7(1)(b) of the AI Act suggests a reverse conclusion: AI systems that pose lower risks to the protected interests mentioned, than the cases already regulated, may not be included. Therefore, the draft specifies an undefined minimum risk level that must be derived by the user from the examples given in Annex III, without specifying the relevant criteria. At the same time, the draft deprives itself of the necessary flexibility for the application of the AI Act to cases that have not yet been identified as sufficiently risky and which, in terms of their risks, are to be located above the threshold for non-regulation but below the risk potential expressed by the use cases listed in Annex III.

\subsection{AI systems with special needs for transparency}
Finally, Art. 52 of the AI Act mentions AI systems that do not pose a high risk in the sense described above, but which nevertheless pose indirect restrictions on the interests of third parties protected by fundamental rights due to their specific context of application. The draft regulation includes three categories of AI systems.

First of all, AI systems intended for interaction with natural persons are covered (para. 1). In this regard, the AI Act demands that such systems inform the natural persons - usually the users of the system - that they are not interacting with a human being, but with an AI system. This requirement does not apply if the fact that the user is interacting with an AI system is obvious due to the circumstances and context of the use of the system, i.e., in such cases where there is no need for corresponding information. An exception is also provided for purposes of law enforcement, insofar as corresponding procedures are permitted by law; in these cases, however, there would be a risk that such measures would lose their effectiveness due to the obligation to provide information. Our investigation has shown, particularly with regard to the application of Art. 52(1), that the term "interaction" is certainly open to interpretation. This challenge is discussed in detail in Section~\ref{sec:discussion}.

Also covered are AI systems for emotion recognition and systems for biometric categorization of persons. In this respect, Paragraph 2 demands that users of such systems inform the persons affected by the system - i.e., those persons whose emotions are recognized or who are biometrically categorized - about the operation of the system. Here, too, exceptions are provided for law enforcement purposes.

Finally, Art. 52(3) of the AI Act obliges users of AI systems that are used to generate so-called deepfakes to disclose the fact that the generated content was artificially created or manipulated. The regulation describes the scope of application as covering AI systems that generate or manipulate image, sound or video content that noticeably resembles real persons, objects, places or other facilities or events and would falsely appear to a person to be real or true.

All cases of Art. 52 AI Act concern AI systems for which transparency and information requirements exist for certain actors due to the type of use. The regulation, thus, establishes a further class of AI systems, namely AI systems with special transparency requirements. The classification into this class is recognizably independent of the classification of an AI system as a high-risk AI system: According to the systematics of the AI Act, the requirements of Art. 52 AI Act also apply to those AI systems that are also to be regarded as high-risk AI systems within the meaning of Art. 6 AI Act. This double counting affected some of the use cases studied.

\subsection{Low-risk AI systems}
Finally, the (initially) last class of AI systems is formed by those AI systems that are not explicitly classified as a separate risk class by the AI Act because they are not intended to be subject to the regulation as a whole. AI systems that cannot be assigned to any of the classes outlined above are therefore implicitly classified as low to no risk AI systems. In evaluating the numerous actual use cases, the question has sometimes arisen as to whether every system not covered by the current status actually poses such low risks that it does not require regulation. This concerned, for example, weather forecasting systems (e.g., ID 494 and ID 495) or Smart Home applications (e.g., ID 656).

These systems are addressed solely in Art. 69 AI Act. The regulation encourages the drawing up of codes of conduct intended to foster the voluntary application to AI systems other than high-risk AI systems (para. 1). Moreover, with regard to such systems, the establishment of further codes of conduct shall also be encouraged and facilitated (para. 2).

\subsection{General purpose AI systems}
\label{sec:multi}

As a new category, the latest draft of the AI Act introduced general purpose AI systems. According to the definition in Art. 3 No. 1b), the term describes AI systems that are intended by the provider to perform generally applicable functions. Examples for general purpose AI systems are, inter alia, image and speech recognition, audio and video generation, pattern detection, question answering and translation. Even trained models fall under the scope of the relevant provisions (see Art. 4a(2) AI Act).

Providers of general purpose AI systems must comply with the general requirements for high-risk AI systems set out in Art. 8 to Art. 15 AI Act. In addition, some of the specific obligations for providers in Art. 16 must be observed in any case, such as the obligation to conduct a conformity assessment, to take necessary corrective actions or to affix the CE marking to the system.

The provisions on general purpose AI systems shall not apply when the provider has explicitly excluded all high-risk uses in the instructions of use or information accompanying the system (Art. 4c(1) AI Act). However, the provider should not be allowed to rely on this exclusion if there are sufficient reasons to consider that the system may be misused (Art. 4c(2) AI Act). A 'misuse' in terms of this rule covers in particular any high-risk use which is excluded by the provider.

\section{Data}
We conducted a non-representative quantitative study of the applicability and impact of the AI Act. The objectives were to identify ambiguities and inaccuracies, to find examples that we believe would be over- or under-regulated, and to determine the approximate proportion of systems that would be regulated under the AI Act and those that do not. For this, we analyzed a data base of AI project descriptions w.r.t. to their most likely classification into the four risk-related groups in the following way.

\subsection{Description}
The study is based on a database of ``AI products, services and development projects in Germany'', collected by the \textit{Plattform lernende Systeme}\footnote{https://www.plattform-lernende-systeme.de/ki-landkarte.html, last accessed on 27.01.2022} (PLS). The \textit{Plattform lernende Systeme} is a project funded by the German ministry for Research and Education, where invited members from industry, science, NGOs and politics can discuss ideas about how to use learning systems in society. Among others, the platform hosts this database in which every company or person can self-describe an AI project they are running. While the self-selection bias cannot be neglected, this is the only dataset we know of that describes some hundred AI projects of one European country. We used this as a basis to understand the proportions in which AI projects in Germany might fall into the four different risk classes. 

We received the data set in the fall off 2019, which comprised 760 cases at that time. Since the existing descriptions were not written for the purpose of analysis regarding the AI Act, many entries could not be meaningfully processed and were sorted out in several preprocessing phases as described below. Each entry in the cleaned data set was then independently reviewed by two computer scientists and one legal scientist to determine which risk class under the AI Act it would fall into and with what reasoning. The results were then compared and discrepancies discussed and, whenever possible, resolved. The results were statistically analyzed and special cases were examined in greater depth.

\section{Method}
Our method consisted of a preprocessing phase in which we cleaned the data, followed by an independent classification procedure by three reviewers, and a mediation phase in which differing classifications for each item were discussed. These phases are described in the following.

\subsection{Preprocessing}
Each entry was assigned an ID, then the order of the entries was randomized. After that, each entry was checked to determine whether or not it could be used in our study. Entries were removed if at least one of the following criteria applied:

\begin{itemize}
    \item The description was empty or contained no information about a product (e.g., only introduction of a company, project or research goal).
    \item The description was not sufficiently specific to understand the product or be able to assess it (e.g., ID 485).
    \item The product did not contain an AI component or it was not identifiable what exactly the AI component is (e.g., ID 662).
    \item The description focused on a technology or research approach without naming specific applications that are necessary for a categorization under the AI Act (e.g., ID 468).\footnote{Note that many of these cases could most likely be classified with regard to the new risk-category of general purpose AI systems (see Section~\ref{sec:multi}) as the intended use of a system is now less relevant due to the inclusion of possible uses in a high-risk context.}
    \item The description listed several systems and applications that would fall into different categories and would therefore need to be examined individually (e.g., ID 121).\footnotemark[\thefootnote])
    \item The entry was a duplication (identical text or the same product, e.g., ID 15 and ID 295).
\end{itemize}
The cleaning was done in multiple iterations, as some problems only emerged when we were discussing specific entries.

For the first iteration, four student research assistants\footnote{Shari Paul, Paula Poepl, Melissa Madadi and Anna-Sophie Schmidt} from legal studies were taught computer science basics in the context of AI and discussed several entries together for preparation. Together with a student research assistant from computer science\footnote{Philipp Bird} they each examined all entries for obvious problems. This concerned 8 cases without any description and 66 cases that didn't focus on a product. The students also noted several cases in which they were not sure whether they can be meaningfully processed. Two doctoral students\footnote{Marc P. Hauer and Tobias D. Krafft} from computer science examined those cases in a second iteration and additionally removed 77. After this step 609 entries remained. 

The remaining entries were divided into packages of 50 entries each, to avoid examining too many entries at once and thus minimize errors due to carelessness.

\subsection{Independent classification by three reviewers}
With these entries, the study was conducted by two doctoral students\footnotemark[\thefootnote]) from computer science and one postdoc\footnote{Dr. Andreas Sesing-Wagenpfeil} in legal science. Within one week, each of us reviewed (the same) 50 entries to determine which paragraph of the AI act would categorize the respective product into which regulatory class. Relevant questions and discussion points were noted. If no tagging was possible due to ambiguities or arising questions regarding the AI Act or technical details, the option \textit{not sure} could be tagged.

\subsection{Mediation phase}
After every 50 entries, we met to compare our results and discuss discrepancies. In this process, a total of 74 more entries were removed, either because the discussion revealed that the description was not sufficient to classify the product according to the AI act (70 cases), or because further duplicates were identified (4 cases). After this step 535 entries remained. In the case of initial disagreement, the discussions usually led to an agreement being reached very quickly. The causes of disagreement were usually problems in the interpretation of legal texts on the side of the computer scientists or an improper understanding of the technical system on the side of the legal scientist.

\subsection{Final classification}
\label{sec:final}
Based on the AI act proposal, some AI systems can potentially be addressed by multiple paragraphs that might categorize them into different classes. Art. 52(1) states that 'Providers shall ensure that AI systems intended to interact with natural persons are designed and developed in such a way that natural persons are informed that they are interacting with an AI system, unless this is obvious from the circumstances and the context of use. [...]'. However, most systems interact with humans in some way, the concept of interaction is not legally defined . Additionally, there is no description under which conditions a system is 'obviously' an AI system. Therefore, for our classification, we have made the decision to always choose the sharpest sword in case of several possible classes (i.e., ban $>$ high risk $>$ transparency obligations) and to indicate the most specific paragraph in case of several possible justifications for the same category (i.e. Art. 6 para. 1 i.c.w. Annex II, Section B, No. 6 regulating vehicles of classes M, N and O $>$ Art. 6 para. 1 i.c.w. Annex II, Section A, No. 1 for the regulation of machines in general). If we had not decided that the sharpest sword is chosen among several possible categories, presumably considerably more cases would fall under Art. 52(1). But then the question of the definition of interaction and when it is obviously an AI system would also lead to significantly more unclear cases.

If an agreement could not be reached quickly, it was usually not possible even after long discussions. These 21 cases could not be categorized due to inaccuracies in the AI act (e.g., the terms 'interaction' and 'safety component'), or because several paragraphs were equally (un)specific. These cases are not considered for the statistical analysis, as they would distort the result, but are discussed in this paper (see section~\ref{sec:discussion}). 514 cases remain for the statistical analysis.

\section{Results}
In the following, we discuss our results, starting with the interrater agreement and followed by the distribution of risk-categories in the dataset.

\subsection{Interrater agreement}
The interrater agreement is a method to enumerate the agreement of assignments. It is used wherever no definite correct assignment is known (in which case quality measures, like sensitivity or specificity are more reasonable choices)~\cite{mandrekar2011measures}. Generally, there are various methods to compute the interrater agreement~\cite{gisev2013interrater}. For our study, we compare the respective decisions for a regulatory class or the \textit{not sure} option on all entries we could agree or resolve our disagreement on. For 311 of the 514 entries, all of us agreed on the criticality level without any discussion (60,5\%). Given the final classifications in regulatory classes, a random assignment would have left us with an interrater agreement of 26\%\footnote{$0.3113^3+0.0759^3+0.6128^3 = 0.2607$}. Considering that the descriptions of the entries were not intended to be evaluated on the basis of the AI Act and were not phrased accordingly, we consider this level of agreement to be very high. Notably, agreement between any two raters was the least between the legal scientist and one of the computer scientists (on 349 entries, 67,9\%), but between the other computer scientist and the legal scientist (on 379 entries, 73,74\%) and the two computer scientists (on 382 entries, 74,32\%), agreement was almost equal high. In 139 cases, at least one of us tagged \textit{not sure} (27,04\%), while in 3 cases all of us tagged \textit{not sure}.

Neither the agreement rate, nor the number of \textit{not sure} tags significantly changed over time with increasing experience. The causes of uncertainties and disagreements were rather rarely due to a lack of experience with the application of the AI Act or the general understanding of AI Systems. More often, the difficulty for the computer scientists lied in deeper understanding of other legal texts, as the AI Act contains many references. The legal scientist, on the other hand, was challenged by understanding very specific AI applications that lacked a proper explanation for non-domain experts. Further problems lied in ambiguous descriptions of entries and definition gaps in the AI Act (see section~\ref{sec:discussion} for extreme examples we could not resolve in our discussions).

\subsection{Distribution of risk-categories in the dataset}
From 514 cases for which we could agree on a risk level, 199 would fall under the new regulation while 315 would be classified as low-risk. Among the regulated cases, none of them would fall under prohibition, 160 would fall under high-risk ($\sim$31,13\% among all), and 39 would fall under transparency obligations only\footnote{As mentioned earlier, the classification as high-risk does not necessarily exclude a system from also falling under the transparency obligations set out in Art. 52 AI Act.} ($\sim$7,59\% among all) (see Table~\ref{tab:classification}). Of the 160 high-risk cases, the categorization of 98 cases is based on Art. 6 para. 1. Annex II, Section A, No. 11 (medical devices) (see Table~\ref{tab:high_risk}\footnote{The table refers to Annex II in the initial draft of the AI Act. In the latest version, civil aviation is now No. 8, whereas No. 7 covers an additional regulation on motor vehicles and associated trailers.}). Of the 39 transparency obligation cases, the categorization of 34 cases is based on Art. 52 para. 1 (interaction with natural persons) (see Table~\ref{tab:transparency}).

\begin{table}[h]
\centering
\caption{}
\label{tab:classification}
\begin{tabular}{l|r r}
\textbf{Number of cases with unambiguous category and reasoning}                & \textbf{514}  & \textbf{(100\%)} \\ \hline
Number of cases that would  be prohibited              & 0 & (0\%)       \\ \hline
Number of cases that would be categorized as high-risk & 160 & (31,13\%) \\ \hline
Number of cases that would be categorized as cases with transparency obligations & 39 & (7,59\%)          \\ \hline
Number of cases that would be categorized as low-risk  & 315 & (61,28\%)
\end{tabular}%
\end{table}

\begin{table}
\caption{Number of use cases categorized as high-risk applications based on Annex II and Annex III in the initial draft of the AI Act.}
\label{tab:high_risk}
\resizebox{\textwidth}{!}{
\begin{tabular}{l|r}
Art. 6 para. 1. Annex II, section A, No. 1 (machinery)  & 16 \\ \hline
Art. 6 para. 1. Annex II, section A, No. 10 (burning gaseous fuels) & 1  \\ \hline
Art. 6 para. 1. Annex II, section A, No. 11 (medical devices) & 98 \\ \hline
Art. 6 para. 1. Annex II, section A, No. 12 (in vitro diagnostic) & 2  \\ \hline
Art. 6 para. 1. Annex II, section B, No. 2 (two- or three-wheel vehicles and quadricycles) & 5  \\ \hline
Art. 6 para. 1. Annex II, section B, No. 4 (marine equipment)  & 3  \\ \hline
Art. 6 para. 1. Annex II, section B, No. 6 (motor vehicles and associated trailers, systems, components, ...) & 9 \\ \hline
Art. 6 para. 1. Annex II, section B, No. 7 (civil aviation) & 1  \\ \hline
Art. 6 para. 2. Annex III, No. 1            (biometric identification and categorization) & 8                      \\ \hline
Art. 6 para. 2. Annex III, No. 2             (critical infrastructure) & 6                      \\ \hline
Art. 6 para. 2. Annex III, No. 3             (education and vocational training) & 1                      \\ \hline
Art. 6 para. 2. Annex III, No. 4             (Employment, workers management and access to self-employment) & 6                      \\ \hline
Art. 6 para. 2. Annex III, No. 5             (private and public services and benefits) & 1                      \\ \hline
Art. 6 para. 2. Annex III, No. 6             (law enforcement) & 2                      \\ \hline
Art. 6 para. 2. Annex III, No. 7             (migration, asylum and border control management) & 1                      \\ \hline
\end{tabular}%
}
\end{table}

\begin{table}[h]
\centering
\caption{Number of use cases categorized as receiving transparency obligations based on Art. 52 in the initial draft of the AI Act.}
\label{tab:transparency}
\begin{tabular}{l|r}
Art. 52 para. 1 (interaction with natural persons)                                    & 34                     \\ \hline
Art. 52 para. 2 (emotion recognition and biometric categorization)                                    & 5                      \\ \hline
Art. 52 para. 3 (generation or manipulation of image, audio or video content)                                    & 0
\end{tabular}%
\end{table}

From the 21 cases that raised questions, for 8 cases it was unclear whether the term \textit{safety component} applies (Art. 6 Abs. 1 i.c.w. Annex II, Abschnitt A und B). For 4 cases, we could not assess whether the product could result in an overload of the power grid (Art. 6 para. 2 i.c.w. Annex III, Nr. 2), for 6 cases the term \textit{interaction} was not sufficiently defined (Art. 52(1)) and 3 cases each raised unique questions.


\section{Discussion of the results}
\label{sec:discussion}
It is not surprising that no harmful AI systems were found in this study: the self-selection bias of how entries came into the database makes it very unlikely that anyone would publicly announce a system that is deemed to be harmful, e.g., lethal autonomous weapons. Most high-risk systems are in the medical sector. This is not surprising, as it was a highly regulated area before. However, no statement about the distribution of AI products can be derived from this, as the collection of use cases considered does not necessarily correspond to their real distribution. The ratio of systems classified as low-risk (61,28\%) to those not classified as low-risk (38,72\%) serves as an indication that there is no need to fear over-regulation by the AI Act. However, it may be that the \textit{Plattform lernende Systeme} just contains an overproportionate number of low-risk systems.

As mentioned before, we were not able to categorize 21 use cases, due to challenges within the AI act. The exact challenges are explained with case IDs below. The full texts of the respective cases are listed in~\ref{ap:use_cases}.

\subsection{Challenges of applying the AI Act}
The regulation refers, especially in the context of the definition of high-risk systems, to 'safety components'. According to Art. 3 No. 14 AI Act, a \enquote{safety component of a product or system means a component of a product or of a system which fulfills a safety function for that product or system or the failure or malfunctioning of which endangers the health and safety of persons or property}. Whether a use case contains a safety component according to this definition could only be roughly estimated on the basis of the description alone. In 8 cases we could not come to a clear conclusion:

\begin{itemize}
    \item Case 262 
    involves robotic systems for decontamination in hostile environments. It could be argued that their failure makes human intervention necessary. Since without this product human intervention would be inevitable, it seems reasonable not to classify this product as a high-risk AI system. However, the wording of the regulation is not so clear.
    \item Case 141
    is about the AI-supported development of sea ice maps to support the navigation of ships. It remains unclear if sea maps are somehow integrated in ships; if so, the failure of this system would reduce the safety of the component of a ship. This case raises the question if a 'safety component' must be integrated in a product.
   \item Case 17 
    introduces a hearing aid whose noise reduction is optimized with the help of AI. A failure might result in severe hearing difficulties as it is not clear from the description whether there is a fall back to traditional noise reduction. However, the distinction between a 'safety function' and the main function of an AI system is not clear at all.
    \item Case 42 
    presents the AI-supported production of plasma. Based on the description we could not asses, whether any mistakes in the production process may lead to safety critical consequences.
    \item Case 210 
    involves the development of AI-based radar systems. Whether or not these are safety components depends heavily on how they are integrated and used in an overall system. The question arises as to whether the manufacturer of an initially unregulated AI component would need to determine a specific intended use in order to determine whether it falls under the AI Act. Thus, the responsibility for compliance under the AI Act could fall on the user in the case of a non-intended use. However, the case of a general purpose radar system would most likely fall under the scope of the general purpose AI systems which have become part of the AI Act in its recent version (see Section~\ref{multi}).
    \item Three cases involve predictive maintenance products (ID 73, ID 367, ID 434). 
    It is unclear whether, if these products fail, classic quality assurance mechanisms will kick in. In this case, these products would presumably not be safety components. Even if this were currently the case, it seems likely that redundant quality assurance mechanisms will be dismantled as trust in predictive maintenance products increases. In this scenario, the products would presumably be safety components. Such considerations are not taken into account by the AI act.
\end{itemize}

According to the heading of Annex III, No. 1, this category of high-risk AI deals with \enquote{Biometric identification and categorization of natural persons}, but the detailed explanation only refers to \enquote{AI systems intended to be used for the ‘real-time’ and ‘post’ remote biometric identification of natural persons}, not considering categorizations at all. Case 682 describes a digital advertising surface whose content adapts to passers-by. The passers-by are categorized on the basis of current video information in order to select suitable advertising content. No data is stored in the process. Due to the inconsistency between heading and description of Art. 6 para. 2 i.c.w. Annex III, No. 1, it remains unclear whether this case should be considered a high-risk product. This issue has been addressed by the latest amendmends to the draft AI Act: The category is now purely described as "biometrics", the (single) use case is not limited to 'post' biometric identification any longer. Nonetheless, the use case is still restricted to 'remote' biometric identification. Thus, Case 682 should now be classified as high-risk AI.

Art. 6 para. 2 i.c.w. Annex III, No. 2 addresses \enquote{AI systems intended to be used as safety components in the management and operation of road traffic and the supply of water, gas, heating and electricity}. This might mean that any AI system that might lead to power supply failure can be considered a safety component. Cases 189, 529, 93, 274 deal with products that allow dynamic production and/or consumption of energy. Whether scenarios that threaten the power supply are realistic cannot be assessed on the basis of the descriptions. In the current version of the draft AI Act, 'critical digital infrastructure' has been added to the use case. We welcome this amendment, as certain use cases dealing with digital infrastructure -- which is highly important for society and our daily lives -- did not clearly fall within the scope of the initial AI Act draft (e.g., ID 679).

As Art. 6 para. 2 i.c.w. Annex III, No. 8 deals with \enquote{AI systems intended to assist a judicial authority in researching and interpreting facts and the law and in applying the law to a concrete set of facts}. That means a categorization might not only depend on the product but also on the operator. Case 482 deals with a system that is able to analyze chats to identify adults that try to \enquote{obtain private information and photos or even arrange personal meetings} with children. If used by a judicial authority this product would most likely fall under this provision. The description did not mention a specific operator though.

Even though we agreed on how to handle the unclear term \textit{interaction} (see Section~\ref{sec:final}), we could not agree on whether the cases 25, 670, 232, 537, 539 an 401 would fall under Art. 52(1).
\\\\
The cases that made it difficult for us to categorize are particularly interesting. They show, on the one hand, where the AI Act needs further sharpening, and on the other hand, how much information is needed regarding the concrete application of a technical system. Additionally, the distribution of the other cases gives an indication of where more regulation is needed and where we could also work with lighter regulation models.

\subsection{Need for interdisciplinarity}
It has become clear that the legal assessment of AI systems based on the AI Act requires both a deep understanding of the law and technical expertise in the AI context. The legal expertise was necessary to get to the bottom of individual legal terms (such as the legal definition of the term \textit{safety component}) and the many references to other legal texts. For many entries this was the reason for the two computer scientists to tag \textit{not sure} (65 and 47 times respectively). The expertise in the field of computer science was necessary to understand what is behind certain technologies. For many entries this was the reason for the legal scientist to tag \textit{not sure} (59 times). Only in 3 cases all participants used the \textit{not sure} tag.


\section{Threats to validity}
This study is based on a shaky ground, as the database entries were input by whoever wanted to add an entry. In principle, the entries can thus be faked or only describe ideas which are not yet implemented. However, each/most entries were accompanied with a link on which further information could be found which induces more effort for fake entries. Additionally, we are not aware of any advantage for adding AI projects to the \textit{Plattform lernende Systeme} other than contributing to the landscape of projects in Germany to foster awareness. 
Due to the self-selection process, the percentage of high-risk or low-risk applications of AI could be higher or lower: it might be that some projects were deemed too unimportant to report or people might not even realize that AI is part of a project\footnote{The reverse also happened: Projects were listed that did not seem to contain a component that would be addressed by the AI act. These were cleaned in the preprocessing step.}. High-risk projects might not want to get public attention. 
Certainly, the database is not known to all persons or companies that use AI systems. Governmental uses of AI systems were not systematically announced here, either. 

Another threat to validity is that the descriptions of the projects were very short. Those, that did not lend itself to a classification, were thus deleted in the preprocessing, however, it might be that a classification would be different if more was known about a specific system and its application.

\section{Summary and future questions}

The percentages of the different classes can only be taken with a large grain of salt. However, the categorization of use cases is as thoroughly worked out and communicated as it can be at the moment. For the \textit{Plattform lernende Systeme} database, the fear of overregulation does not seem to be confirmed. Furthermore, the high-risk systems in our study were mostly medical systems, which already belong to the most regulated systems in Germany, i.e., they pose a well-known risk and are regulated accordingly. Nonetheless, the public observation of AI based systems on the media seems to focus much more on the high-risk systems. The method can be transferred to other databases to get a more complete picture. It can also be repeated to get a conclusion on the final version of the AI Act when it enters into force. The procedure can also be helpful in the context of the regular review of the list of AI systems in Annex III by the European Commission (see Art. 84 para. 1b AI Act) in order to determine the need to add further AI systems to Annex III (or, if necessary, to delete some).

Our study identified several formulations in the AI Act that are too imprecise to be able to assess all applications. In addition, a few weaknesses regarding its applicability became apparent. In particular, if an application is to be used in several contexts, the application of the AI Act remained unclear due to the fact that a system or a trained model might be used for different purposes. This problem seems to be significantly reduced due to the integration of general purpose AI systems in Art. 4a--4c of the recent draft of the AI Act. Developers of AI systems will have to investigate the possible scope of the systems they provide.

The AI Act is expected to be passed this year (2023) and then undergo a two-year review period. During this time, the approach presented here could be repeated with the then provisionally final version of the AI Act to assess its impact and identify possible improvement.

\section{Acknowledgement}
The research was performed within the project GOAL \enquote{Governance of and by algorithms} (Funding code 01IS19020; \href{https://goal-projekt.de/en/}{https://goal-projekt.de/en/}) which is funded by the German Federal Ministry of Education and Research and the projects \enquote{Deciding about, by, and together with algorithmic decision making systems} and \enquote{Explainable Intelligent Systems} (Funding code AZ 98 511; \href{https://www.eis.science}{https://www.eis.science}), both funded by the Volkswagen Foundation. 


\bibliographystyle{elsarticle-num}
\bibliography{main}







\appendix
\section{Full text use cases}
\label{ap:use_cases}
All use cases we refer to in this document are listed in full text below. Their original sources are referenced in footnotes, however, the descriptions may change there over time. The entries of finished projects or products that are not available anymore may be removed. The full list of original use cases and how we evaluated them has been uploaded to the journal to provide replicability.

\begin{itemize}
    \item ID 15\footnote{\url{https://www.plattform-lernende-systeme.de/anwendung.html?AID=516}}: Im Projekt TraMeExCo werden am Beispiel zweier klinischer Fragestellungen (Pathologie, Schmerzanalyse), neue Verfahren des Maschinellen Lernens erforscht und an verschiedenen Datensätzen untersucht (Mikroskopiebilder, Schmerzvideos, EKG-Signale). Im Maschinellen Lernen kann grob zwischen zwei Arten von Lernverfahren (\enquote{Black-Box}, \enquote{White-Box}) unterschieden werden. Ein \enquote{Black-Box}-Lernverfahren (z. B. Deep Learning) \enquote{lernt} durch Beispiele (Bilder, Töne, Daten, etc.) verschiedene Muster (Klassen). \enquote{Black-Box}-Verfahren werden in den letzten Jahren immer erfolgreicher eingesetzt, allerdings haben sie den Nachteil, dass die interne Verarbeitung der Eingabeparameter für den Betrachter nicht nachvollziehbar ist. Bei einem \enquote{White-Box}-Verfahren kann der Beobachter nachvollziehen bzw. begründen, wie die Eingabedaten verarbeitet werden, allerdings sind diese Verfahren bei weitem nicht so erfolgreich in der Mustererkennung. Die Hauptaufgabe im Projekt ist, durch die Kombination von \enquote{Black-Box}-Lernverfahren mit \enquote{White-Box}-Lernverfahren, die Nachvollziehbarkeit der gefundenen Erkenntnisse für den Arzt zu erhöhen und die erfolgreiche Mustererkennung weiter zu ermöglichen. Hierbei sollen auch die zugrundeliegenden Unsicherheiten von System und Daten beachtet werden. Als Ergebnis des Projektes entstehen zwei prototypische \enquote{Transparente Begleiter für Medizinische Anwendungen}, die die relevanten Informationen in den großen Datenmengen für die Ärzte einfacher zugänglich machen.

    \item ID 17\footnote{\url{https://www.plattform-lernende-systeme.de/anwendung.html?AID=831}}: Die Reduzierung unerwünschter Umgebungsgeräusche ist ein wichtiges Merkmal heutiger Hörgeräte. Daher ist die Lärmreduzierung heute in fast jedem handelsüblichen Gerät enthalten. Die Mehrheit dieser Algorithmen beschränkt sich jedoch auf die Reduzierung von stationären Geräuschen. Aufgrund der Vielzahl unterschiedlicher Hintergrundgeräusche im Alltag ist es schwierig, heuristische Lösungen für alle Geräuschkulissen zu finden. Deep learning basierte Algorithmen stellen eine mögliche Lösung für dieses Dilemma dar, aber manchmal fehlt es ihnen an Robustheit und Anwendbarkeit im strengen Kontext von Hörgeräten.

    In diesem Projekt untersuchen wir mehrere deep learning Methoden zur Störgeräuschreduzierung unter den Bedingungen moderner Hörgeräte. Dies beinhaltet eine Signalverarbeitung mit geringer Latenzzeit sowie den Einsatz einer im Hörgerät eingesetzten Filterbank. Ein weiteres wichtiges Ziel ist die Robustheit der entwickelten Methoden. Daher werden die Verfahren mit reale Rauschsignale getestet, die mit Hörgeräten aufgenommen wurden.
    
    Die Reduzierung unerwünschter Umgebungsgeräusche ist ein wichtiges Merkmal heutiger Hörgeräte. Daher ist die Lärmreduzierung heute in fast jedem handelsüblichen Gerät enthalten. Die Mehrheit dieser Algorithmen beschränkt sich jedoch auf die Reduzierung von stationären Geräuschen. Aufgrund der Vielzahl unterschiedlicher Hintergrundgeräusche im Alltag ist es schwierig, heuristische Lösungen für alle Geräuschkulissen zu finden. Deep learning basierte Algorithmen stellen eine mögliche Lösung für dieses Dilemma dar, aber manchmal fehlt es ihnen an Robustheit und Anwendbarkeit im strengen Kontext von Hörgeräten.

    \item ID 25\footnote{\url{https://www.plattform-lernende-systeme.de/best-practice.html?AID=814}}: Elf Milliarden Mahlzeiten werden jährlich in europäischen Kantinen serviert. Die manuelle Bezahlung dieser Mahlzeiten führt nicht nur zu langen Wartezeiten für die Kunden, sondern auch zu hohen Kosten für die Betreiber. auvisus ermöglicht einen Self-Checkout durch Bildverarbeitung mit künstlicher Intelligenz: Kunden platzieren ihr Tablett einfach unter einer Kamera - dort erkennt ein Algorithmus die Gerichte: sofort und präzise.

    \item ID 42\footnote{This use case is no longer available online}: Die Forschungsgruppe Laser- und Plasmatechnologie ist ein Teil der Fakultät Naturwissenschaft und Technik der Hochschule für angewandte Wissenschaft und Kunst Hildesheim/ Holzminden/ Göttingen (HAWK). Die Kernkompetenz der Forschungsgruppe ist die Forschung an kalten Atmosphärendruckplasmen. Im Zuge dessen, werden in der Forschungsgruppe seit vielen Jahren speziell ausgerichtete Plasmaquellen gebaut, erforscht und optimiert.
    
    Da die Plasmaqualität von vielen Parametern abhängt, ist Prozesssicherheit bei Plasmabehandlungen ein wichtiges Thema. Insbesondere Umwelteinflüsse wie Luftdruck, Luftfeuchte und Lufttemperatur haben einen sehr starken Einfluss auf das Plasma. Aber auch Parameter wie der Abstand von Plasmaelektrode zu Substrat oder das Prozessgasgemisch sind von großer Bedeutung.
    
    Um diese und weitere Einflüsse bei der Plasmaerzeugung zu kompensieren, werden in der Forschungsgruppe Sensoren zur schnellen Detektion von Plasmaeigenschaften entwickelt. Da Plasma ein sehr starker elektromagnetischer Störer ist, ist diese Arbeit nicht trivial. Auf Basis der Sensordaten, entwickelt die Forschungsgruppe intelligente Algorithmen, welche Plasmaquellen dauerhaft in einem Arbeitspunkt ausregeln können.
    
    Abhängig vom gewählten Arbeitspunkt können Plasmaquellen unterschiedliche Effekte erzeugen. So können Sie zur Aktivierung, zur Beschichtung oder zur Feinstreinigung von Oberflächen genutzt werden. Aber auch weitere Anwendungen sind in unterschiedlichen Arbeitspunkten möglich. Dazu zählen die Desinfektion von Oberflächen, eine erhöhte Keimung von Saatgut, die Beseitigung von Schädlingen und das Heilen von Wunden.
    
    Durch die Forschung an intelligenten Algorithmen, können Plasmaquellen geschaffen werden, die selbstständig vorgegebene Arbeitspunkte anfahren und halten. Auf Basis dieser Technologie kann Prozesssicherheit für vielfältige Plasmaanwendungen dauerhaft und unabhängig von Umwelteinflüssen gewährleistet werden.

    \item ID 73\footnote{\url{https://www.plattform-lernende-systeme.de/anwendung.html?AID=765}}: Die Flugzeugproduktion steht vor zahlreichen technischen Herausforderungen, wie z.B. großen Produktabmessungen, komplexen Fügeverfahren und der Organisation von Montageaufgaben im Rahmen der stetig steigenden Flugzeugproduktion.
    
    Durch die Ausstattung eines Human-Robot-Collaboration-fähigen Systems mit Kamera, Kraft-Drehmoment- und Laser-Linien-Sensoren ist eine kollaborative Qualitätsprüfaufgabe von Mensch und Roboter möglich. Zusätzlich werden die Sensoren eingesetzt, um den Nietvorgang in Echtzeit zu überwachen und Endkontrollaufgaben mittels KI durchzuführen, was den Produktionsprozess schneller und effizienter macht.
    
    Die aus dem Scan gewonnenen Daten werden durch ein trainiertes neuronales Netzwerk geleitet, um das betreffende Objekt als gut oder schlecht in Bezug auf die Qualität zu klassifizieren. 
    
    Um eine umfassende Lösung zu entwickeln, visualisiert eine Augmented-Reality-Anwendung den aktuellen Status des Prozesses und ermöglicht eine intuitive Interaktion mit dem System. Dies ermöglicht es dem Bediener, notwendige Wartungsarbeiten mühelos durchzuführen.

    \item ID 93\footnote{This use case is no longer available online}: Der steigende Anteil energieproduzierender Privatverbraucher (sog. Prosumer), die ihren eigenen Strom aus erneuerbaren Energien mit hohen Schwankungen produzieren, verlangt von den Energieversorgern nachvollziehbare Vorhersagen und ein viel genaueres Verständnis der Entwicklung des lokalen Energiebedarfs als bislang möglich, um diesen mit der lokalen Produktion zu harmonisieren und optimal zu gestalten. Diese Situation stellt eine besondere Herausforderung für kleinere Stadtwerke und Energieanbieter dar, die keinen Zugang zu sophistizierten Modellierungs- und Prognosemethoden und zum für ihre Anwendung erforderlichen Fachwissen haben. Auch stellen die oft nicht direkt nachvollziehbaren Funktionsweisen und Ergebnisse der zugrundeliegenden KI-Methoden existierender Lösungsansätze eine große Hürde für ihre Anwendung dar. 
    
    Diesem Problem wird im SIT4Energy-Projekt durch einen Ansatz begegnet, welcher Methoden des maschinellen Lernens mit visueller Analytik kombiniert, um erklärbare Vorhersagen und Analysen des lokalen Energiebedarfs in Prosumer-Umgebungen zu ermöglichen. Auf dieser Basis wird ein nutzerzentriertes interaktives Werkzeug entwickelt, mit welchen auch kleinere Energieversorger effektive Vorhersagen des lokalen Energiebedarfs anhand historischer Daten und Simulationsparameter für verschiedene Szenarien auf eine für sie nachvollziehbare Weise erstellen und analysieren können.

    \item ID 121\footnote{This use case is no longer available online}: BirdieMatch ist das Job-Matching-Portal für Logistiker und funktioniert nach dem Prinzip von Partnerbörsen. Der Rekrutierungsprozess wird durch Matching-Technologie digitalisiert. So ist eine innovative Online-Plattform entstanden, die als digitaler Headhunter fungiert. Das Besondere ist neben der Technologie die Spezialisierung auf Logistikberufe. Mehr als 600 logistikspezifische Matching-Kriterien bringen Stellen und Arbeitskräfte auf fachlich/sachlicher Ebene schnell und einfach zusammen. Zudem werden die Heart-Skills der Kandidaten sowie Benefits der Unternehmen einbezogen. Die Matching-Technologie ist nicht nur durch mehr als 20-jährige Erfahrung im Bereich Logistik \& Recruiting entstanden, sondern wird auch ständig durch neue Trainingsdaten weiter optimiert. 
    
    BirdieMatch konzentriert sich derzeit allein auf Logistiker und zeichnet sich hier durch die Möglichkeit sehr tiefreichender und fachlicher Beschreibungen von Kenntnissen, Erfahrungen und Jobanforderungen aus, die miteinander gematcht werden. 

    \item ID 141\footnote{\url{https://www.plattform-lernende-systeme.de/anwendung.html?AID=687}}: Mittels neuronaler Netzwerke werden Signaturen von verschiedenen Meereisklassen auf SAR Bildern vollautomatisch unterschieden und so hochauflösende Eiskarten in nahe-Echtzeit erstellt. Darüber hinaus wird die Eisdrift abgeleitet und so eine Kurzzeitvorhersage der Eisverhältnisse ermöglicht. Auf Basis dieser Daten werden Risikokarten erstellt, die zur Navigationsunterstützung verwendet werden können oder auf deren Basis direkt Routenempfehlungen gegeben werden. Besonders für Schifffahrtswege wie die arktische Nord-Ost- und Nord-West-Passagen mit hoher wirtschaftlicher Attraktivität aufgrund der kürzeren Distanz wird so die Sicherheit des Schiffsverkehrs erhöht.  

    \item ID 189\footnote{\url{https://www.plattform-lernende-systeme.de/anwendung.html?AID=642}}: Um die Energiewende zu schaffen, bedarf es einer Vielzahl von Anstrengungen. Eine ist die Ausnutzung des Lastenverschiebungspotenzials, welches durch den Einsatz von erneuerbaren Energien immer mehr an Bedeutung gewinnt. Dies bedingt eine Veränderung und Steuerung des Verhaltens von Energieerzeugern und -verbrauchern. Die hierfür notwendigen Anreize können z.B. über dynamische Stromtarife gesetzt werden. Auf der Basis von Verbrauchsinformationen, wie sie bspw. von Smart-Metern zur Verfügung gestellt werden, eröffnen sich im Zusammenspiel mit Heimautomationsanlagen neue Möglichkeiten für eine Lastenverschiebung im privaten Umfeld. Um die Kundenakzeptanz zu erhöhen, muss sowohl die Auswertung der Verbrauchsinformation als auch die Auswahl des Tarifs automatisch erfolgen können.
    
    Eine automatisierte Ausschöpfung des Lastenverschiebungspotenzials kann nur erfolgreich sein, wenn Anreize gesetzt werden (hier durch dynamische Stromtarife), und wenn auch technisch wenig interessierte Nutzer diese verwenden können. Daher kommen auf Verbraucherseite maschinelle Lernalgorithmen zum Einsatz, die ohne aktive Eingriffe des Nutzers den Stromverbrauch optimieren und sich an Veränderungen anpassen können, ohne für den Nutzer negative Entscheidungen zu treffen.

    \item ID 210\footnote{\url{https://www.plattform-lernende-systeme.de/anwendung.html?AID=1252}}: Die hochauflösenden automotive 3D Radarsensoren der neuesten Generation der Astyx GmbH sind erstmalig in der Lage, genügend dichte Daten zu liefern, anhand derer eine robuste 3D Objekterkennung mit geringer Anzahl an Falschzielen und somit stark erhöhter Zuverlässigkeit möglich ist.
    
    Diese robuste Erkennung kann jedoch nur mit Deep Learning basierten, auf untrainierte Situationen generalisierenden Verfahren der KI bewerkstelligt werden.
    
    Dadurch erlangt die Astyx GmbH erstmalig die algorithmische Befähigung eines solchen, hochauflösenden Automotive Radars zur wetterunabhängigen und störsicheren Fahrumfelderkennung für die Anwendung im automatisierten bis autonomen Fahren (Level 3-5).

    \item ID 232\footnote{This use case is no longer available online}: Um Kunden einen besseren Überblick ihrer Finanzen zu bieten hat die Commerzbank AG gemeinsam mit Accenture reagiert und das digitale, KI-basierte Produkt \enquote{CashRadar} in einem hochgradigen agilen und vollständig Cloud-basierten Liefermodell entwickelt. Herzstück der Applikation ist der Prognose-Algorithmus basierend auf der Technologie \enquote{R}.
    
    Mithilfe des CashRadars können Unternehmer ihre zukünftige Liquidität berechnen. Basierend auf den Umsatzdaten des Nutzers berechnet der CashRadar eine Liquiditätsprognose von bis zu 120 Tagen in die Zukunft. Der Liquiditätsmanager stellt dem Nutzer eine Gesamtübersicht der gebuchten und prognostizierten Zahlungsein- und -ausgänge, über einen Zeitraum von insgesamt 7 Monaten, bereit. 
    
    Mit CashRadar behält der Unternehmer alle Umsätze im Blick, kann Handlungsbedarf frühzeitig identifizieren und kennt die aktuellen Möglichkeiten und zukünftigen Potentiale seines Unternehmens.

    \item ID 262\footnote{This use case is no longer available online}: Müssen chemisch verseuchte Areale saniert oder kerntechnische Anlagen zurückgebaut werden, sind die Arbeiter – allen Vorsichtsmaßnahmen und Schutzausrüstungen zum Trotz – erheblichen Gesundheitsrisiken ausgesetzt.
    
    ROBDEKON steht für \enquote{Robotersysteme für die Dekontamination in menschenfeindlichen Umgebungen} und ist der Erforschung von autonomen oder teilautonomen Robotersystemen gewidmet, damit Menschen der Gefahrenzone in Zukunft fernbleiben können.
    
    Das FZI erforscht im Rahmen des Projektes, wie mobile Roboter mit künstlicher Intelligenz beispielsweise automatisierte kraftbasierte Greifstrategien beim Bergen von Gefahrenstoffen nutzen und so den Bediener entlasten. Für die weitere Nutzerunterstützung mit Hilfe künstlicher Intelligenz arbeitet das FZI außerdem an den Themen 3D-Umgebungserfassung, sichere Navigation, aber insbesondere auch an intuitiven Eingabemethoden mit Augmented- oder auch Virtual-Reality-Technologien.

    \item ID 273\footnote{\url{https://www.plattform-lernende-systeme.de/anwendung.html?AID=542}}: Die automatische Online-Klassifikation kritischer Netzsituationen (z.B. Generator- und Leitungsausfälle) auf Basis hochdynamischer Phasormessungen ermöglicht die Einleitung von Gegenmaßnahmen zur Vermeidung von Versorgungsausfällen und erhöht die Netzstabilität. Hierzu wurde neuartige Assistenzsysteme entwickelt zur Ergänzung klassicher Übertragungsnetzleitwarten unter Einsatz maschineller Lernverfahren (z.B. Rekurrente Neuronale Netze). Verschiedene Ansätze wurden hierzu umgesetzt zur Auswertung von Online-Daten sowie zur unüberwachten Extraktion kritischer Fehlermuster in historischen Massendaten. Die Umsetzung der Methoden erfolgte in Python sowie Matlab unter Verwendung verschiedener ML-Toolboxen (z.B. Keras, Scikit, Tslearn).

    \item ID 274\footnote{This use case is no longer available online}: Verluste des Höchstspannungsnetzes müssen vom Übertragungsnetzbetreiber (ÜNB) an der Börse im Voraus beschafft werden. Dazu wird ein separater Bilanzkreis vom ÜNB bewirtschaftet. Grundlage der Beschaffung ist eine Prognosemethode, die vom Fraunhofer IOSB-AST entwickelt wurde und auf KI\_Technologie basiert. Die Methode ist Teil der Modellbibliothek der Softwarelösung EMS-EDM PROPHET(r). Weiterhin ist die Methode in als Deep Learning Methode in Keras umgesetzt.

    \item ID 295\footnote{\url{https://www.plattform-lernende-systeme.de/anwendung.html?AID=516}}: Ziel dieses Projekts ist die Erforschung und Entwicklung geeigneter Methoden zum robusten und erklärbaren Maschinellen Lernen auf zwei komplementären Anwendungen aus dem Gesundheitsbereich. Dazu soll der Nachweis der Möglichkeit einer (diagnostischen) Vorhersage auf komplexen Daten und der damit verbundenen Transparenz und Erklärung dieser Entscheidung gegenüber dem klinischen Fachpersonal mittels geeigneten Verfahren erbracht werden. Ein weiteres Ziel ist durch die Modellierung von Unsicherheiten in den Eingangsdaten eine Fehlerabschätzung der Vorhersagen zu ermitteln. Als Fallbespiele werden dabei zwei Szenarien der klinischen Routineversorgung adressiert (Pathologie, Schmerzanalyse). Als Ergebnis entstehen zwei prototypische \enquote{Transparente Begleiter für Medizinische Anwendungen} für diese beiden klinischen Fragestellungen.
    
    Partner im Verbundsprojekt ist die Otto-Friedrich-Universität Bamberg.

    \item ID 367\footnote{\url{https://www.plattform-lernende-systeme.de/anwendung.html?AID=465}}: Das Straßennetz der DACH-Länder unterliegt einem permanenten Alterungsprozess und benötigt eine möglichst lückenlose Zustandserfassung und -bewertung um notwendige bauliche Maßnahmen frühzeitig durchführen zu können. Dazu ist eine regelmäßige Erfassung der Fahrbahnoberfläche notwendig. Bei der bildhaften Erfassung mit Messfahrzeugen wird bereits ein hoher Automatisierungsgrad erreicht. Die Auswertung des Bildmaterials erfolgt jedoch bisher nur durch menschliche Experten. Dieser Prozess ist zeitintensiv und fehleranfällig. Im Forschungsprojekt kann gezeigt werden, dass unter Einsatz von Convolutional Neural Networks vielversprechende automatisierte Erkennungsleistungen erreicht werden. Zur Erhöhung der Generalisierungsfähigkeit soll eine breitere Datenbasis eingesetzt werden, welche unterschiedliche Straßenoberflächen im D-A-CH Raum abdeckt. Dabei soll mittels Maschineller Lernverfahren gezielt untersucht werden, welche Daten den größten Beitrag zu guter Generalisierung leisten.

    \item ID 401\footnote{\url{https://www.plattform-lernende-systeme.de/anwendung.html?AID=431}}: ARADIN ist ein Robo-Advisor zur effizienten Beratung von Privatkunden für Ihre finanzielle Risikoabsicherung und Vorsorge. Die vielfältigen Möglichkeiten einer Absicherung überfordert Berater als auch eine regelbasierte Ermittlung. Basierend auf der Finanzanalyse nach DIN7723 wird gemeinsam mit dem Kunden und Vermittler die KI mit relevanten Daten beliefert.  ARADIN wird aktuell mit Millionen von Fällen trainiert, um optimierte Handlungsempfehlungen für Absicherung und Vorsorge zu ermitteln. Ziel: Steigerung der Beratungsqualität für den Kunden und effizienterer Vertrieb. Das Projekt ist gefördert durch das Bundesministerium für Wirtschaft und Energie auf Beschluss des Deutschen Bundestages.

    \item ID 414\footnote{\url{https://www.plattform-lernende-systeme.de/anwendung.html?AID=423}}: AnyPPA, kurz für Anonymous Predictive People Analytics, beschäftigt sich mit der datenschutzkonformen Auswertung von Daten. Kernziel ist dabei die Verringerung von Defiziten in der Personalentwicklung und -auswahl in Unternehmen, beispielsweise hinsichtlich Diversität und Fairness.
    
    Der Fokus liegt auf fortgeschrittenen prädiktiven Datenverarbeitungsverfahren, mit welchen Daten aus einer Vielzahl von IT-Systemen für die Verbesserung von unternehmerischen Entscheidungen genutzt werden können. anacision denkt \enquote{People Analytics} (die Analyse von Beschäftigtendaten) neu und aus Sicht der Beschäftigten. Zudem werden systematische Schwachstellen in der gesamten Personalentwicklung aufgedeckt und minimiert (z.B. Subjektivität, Biases in Personalauswahl und -entwicklung, Förderung von Diversität und Fairness). Im Ergebnis sollen HR-Verantwortliche dabei unterstützt werden, Entscheidungen zu treffen, die auf empirischen Erkenntnissen basieren und zum Unternehmenserfolg beitragen.

    \item ID 434\footnote{\url{https://www.plattform-lernende-systeme.de/anwendung.html?AID=333}}: Um Wasserverluste in einem Leitungsnetz zu reduzieren, müssen auftretende Schäden zeitnah lokalisiert und das Netz im Sinne einer nachhaltigen Rehabilitationsstrategie kontinuierlich erneuert werden. Die Lokalisierung von auftretenden Schäden ist dabei von verschiedenen Faktoren abhängig. Bei den Stadtwerken München ist es bisher notwendig, das Leitungsnetz kontinuierlich zu  begehen und abzuhorchen, um Schäden zu identifizieren.
    
    Aktuell setzen die Stadtwerke München ein Konzept der bilanzierbaren Zonen um. Ziel der SWM Infrastruktur GmbH\&CoKG ist es, durch den Einbau verschiedener Sensoren im Leitungsnetz Daten zu Durchfluss, Druck, Temperatur und Geräusch aufzuzeichnen, um diese Daten nach vorgegebenen Mustern automatisch durchsuchen zu lassen. Im Ergebnis wird erwartet, dass auftretende Schäden automatisiert und deutlich schneller erkannt werden können. Somit wird die Lecklaufzeit, die maßgeblich für die Höhe der Wasserverluste ist, deutlich verkürzt und das vorhandene Personal kann zielgerichteter eingesetzt werden. 

    \item ID 468\footnote{\url{https://www.plattform-lernende-systeme.de/anwendung.html?AID=288}}: Luminovo entwickelt KI basierte Lösungen zur Automatisierung von repetitiven und kostenintensiven Aufgaben. Das Hybride System unterstützt den Mensch beim Klassifizieren und Überprüfen von Bildern und Texten sowie dem Extrahieren von Informationen. Das System wird parallel zum Menschen in den Prozess integriert und versucht diesen nach und nach besser zu verstehen und zu spiegeln. 

    \item ID 482\footnote{This use case is no longer available online}: Bei der Kommunikation im Internet sind Kinder vielen Gefahren ausgesetzt. Sogenannte Cyber-Groomer schreiben Kinder in Chats an und versuchen, sich ihr Vertrauen zu erschleichen. Damit wird es ihnen möglich, an private Informationen und Fotos zu gelangen oder sogar persönliche Treffen zu vereinbaren. Mithilfe maschineller Lernverfahren und Natural Language Processing analysiert Privalino den Schreibstil eines Chatnutzers und extrahiert linguistische Merkmale. Mit intelligenten Algorithmen kann Privalino typische Muster für Cyber-Grooming lernen und klassifizieren. Erkennt Privalino bei einem Abgleich des analysierten Profils mit denen in der Datenbank hinterlegten Cyber-Grooming-Profilen starke Übereinstimmungen, wird das betroffene Kind gewarnt. Damit leistet Privalino einen Beitrag, das Chatten im Internet für Kinder sicherer zu machen.

    \item ID 485\footnote{\url{https://www.plattform-lernende-systeme.de/anwendung.html?AID=270}}: Die unüberwacht-selbstlernende Predictive Intelligence deckt mit ihren Analysen, Prognosen und Steuerungslösungen Datenmuster in hochkomplexen Prozessen und dynamischen Datenstrukturen auf. Sie ermöglicht im Bereich Industrie 4.0/Smart Production Qualitätsoptimierung und Ausschussreduzierung, vorausschauende Wartung, Energiedispatching und -handel, Anlagensteuerung, Kapazitätsplanung, Logistikoptimierung und Prozesseffizienz. Bei Smart Services sorgt sie für eine bedarfsgerechte Planung, optimalen Ressourceneinsatz, Kommunikationsanalysen sowie optimierte Vertriebs- und Serviceprozesse. Im Bereich Smart Grid ermöglicht es die Lösung, das Potenzial der erneuerbare Energien voll ausnutzen sowie Energieeinkauf \& -verkauf (auch für erneuerbare Energien) präzise durchführen und den Energiehandel vorausschauend zu automatisieren. Im Bereich Smart Building bietet sie die Möglichkeit zu einer vorausschauenden und adaptiven Gebäude-Steuerung.

    \item ID 529\footnote{\url{https://www.plattform-lernende-systeme.de/anwendung.html?AID=182}}: Die Kosten eines großen industriellen Stromverbrauchers hängen von der Gesamtmenge des bezogenen Stroms und der zeitlichen Verteilung des Stromverbrauchs ab. Lastmanagement durch zeitlich flexiblen Einsatz von Maschinen und Geräten verspricht große Einsparpotenziale. Das an der Friedrich-Alexander-Universität Erlangen-Nürnberg (FAU) entwickelte Verfahren basiert auf mathematischer Optimierung und ermöglicht wesentliche Einsparungen bei den lastabhängigen Stromkosten und im Gesamtstromverbrauch im elektrisch betriebenen Schienenverkehr. Durch das Vermeiden zu vieler gleichzeitiger Zugabfahrten kann die im Bahnstromnetz entstehende Last gleichmäßiger über die Zeit verteilt werden. Zusätzlich senkt eine möglichst energiesparende Fahrweise der Züge den Energieverbrauch deutlich. Entscheidend ist, Fahrzeitpuffer optimal zu nutzen, um den Zügen möglichst lange Ausrollphasen zu ermöglichen – dies abgestimmt für alle im Netz verkehrenden Züge. Das entwickelte KI-Verfahren berechnet die optimalen Abfahrtszeitpunkte und Geschwindigkeitsprofile der Züge auf der Strecke und minimiert so Spitzenlast und Energieverbrauch für das Gesamtsystem. Die im Projekt entwickelte Planungssoftware ist übertragbar auf Industrieunternehmen und kann auch dort für die Optimierung des Lastmanagements genutzt werden. 

    \item ID 537\footnote{\url{https://www.plattform-lernende-systeme.de/anwendung.html?AID=50}}: Anbieter von Internetservices müssen Endnutzern individuell die für ihren Anschluss üblicherweise zur Verfügung stehenden Datenübertragungsraten von Internetprodukten bekannt geben. Das schreibt die TK-Transparenzverordnung der Bundesnetzagentur vor. Gerade im Bereich DSL gestaltet sich dies schwierig, da für die historisch gewachsene, maßgeblich auf Kupferdoppeladern basierende TK-Infrastruktur vielfältige Störeinflüsse berücksichtigt werden müssen. Über einen KI-basierten Ansatz lassen sich die funktionalen Abhängigkeiten auf Basis historischer Daten automatisiert erfassen und daraus ein datengetriebenes Prognosemodell ableiten. Durch Berücksichtigung der gesamten Systemstrecke – vom heimischen DSL-Modem über die Teilnehmeranschlussleitung bis hin zur aktiven Technik in den Hauptverteilern – sind exakte Vorhersagen zu erwarteten Übertragungsraten möglich. Stehen hoch aufgelöste Daten im Tages- und saisonalen Verlauf zur Verfügung, kann die Prognose sogar zeitabhängig präzisiert werden. TK-Anbieter können ihre Breitband-Produkte damit für Retail- und Wholesale-Kunden individuell zuschneiden und mit signifikant geringerem Betriebsrisiko anbieten. BTC arbeitet seit 2017 im Auftrag eines Telekommunikationsanbieters daran, ein derartiges KI-gestütztes Prognosewerkzeug zu entwickeln und in den produktiven Betrieb zu integrieren.

    \item ID 539\footnote{\url{https://www.plattform-lernende-systeme.de/anwendung.html?AID=48}}: Das Potenzial semantischer Technologien als Basis für Unternehmensgedächtnisse zeigen – mit diesem Ziel nutzt enviaM das vom DFKI entwickelte wissensbasierte System CoMem/Semantic Desktop. Basis des Piloten für die Abteilung Liegenschaft ist ein Wissensgraph, der aus unterschiedlichen Quellen aufgebaut wurde (z.B. Excel-Dateien, Team-Laufwerke, Dokument- und Bildsammlungen, Geo-Informationen, Bestandssysteme). Darüber hinaus wurde ein Vokabular für die Domäne der Liegenschaften semi-automatisch erstellt. Grundlage waren u.a. die nativen Strukturen des Dateisystems, um Konzepte und deren Beziehungen zu lernen. Wissensdienste ermöglichen nun verschiedene Anwendungsfälle – etwa die semantische Analyse von Dokumenten, die Realisierung einer facettierten Suche oder die proaktive Bereitstellung von Informationen zu einem Kontext. Dieser wird in einem Dashboard für eine Liegenschaft aus dem Wissensgraphen aufgebaut. Unterschiedliche Quellen stellen Informationen bereit und sind unmittelbar nutzbar, etwa die Bestandsakte der Liegenschaft, Katasterinformationen aus dem Liegenschaftssystem, Buchwerte aus dem ERP oder Dokumente aus den Netzlaufwerken.  Der Wissensgraph kann in der täglichen Arbeit gewinnbringend eingesetzt werden, z.B. durch Analyse einer in MS Outlook selektierten E-Mail, der Bereitstellung von Informationen aus dem Wissensgraphen und der Möglichkeit, direkt dort die Recherche zu beginnen.

    \item ID 655\footnote{\url{https://www.plattform-lernende-systeme.de/anwendung.html?AID=115}}: Ziel des Projekts INTUITEL ist es, klassische Learning Management Systeme (LMS) mit einigen Eigenschaften und Fähigkeiten eines menschlichen Tutors auszustatten. Um einen ganzheitlichen Blick auf die jeweilige Lernsituation zu erhalten, werden die bereits in LMS vorhandenen Daten über die Lernenden und den Lernstoff mit zusätzlichen Informationen angereichert. Dazu zählen Daten zum aktuell genutzten Computer, das aktuelle Lernumfeld und vor allem auch allgemeine und themenspezifische pädagogische Modelle. Daraus generiert der intelligente Tutor für jeden einzelnen Lernenden individuelle Lernempfehlungen, die ihn optimal durch den Lernstoff führen.

    \item ID 656\footnote{\url{https://www.plattform-lernende-systeme.de/anwendung.html?AID=114}}: Moderne Gebäude verfügen über komplexe Energiesysteme, bestehend aus verschiedenen Kompenenten wie Heizung, Kühlung und Lüftung. Diese werden oft unabhängig voneinander gesteuert, obwohl mit ganzheitlichen Ansätzen wesentlich höhere Einsparungen erzielbar wären. Das Fraunhofer IIS/EAS arbeitet an intelligenten Systemen zur teil- bzw. vollautomatisierten Betriebsoptimierung sowie intelligenten Algorithmen zur Betriebsdatenanalyse – dem energetischen Alarming. Spezielle Monitoringverfahren ermöglichen hierbei eine schnelle Übersicht über den Gesamtzustand eines Gebäudes sowie auch den detaillierten Einblick in die Ursachen eines Fehlverhaltens. Ein weiterer Schwerpunkt liegt in der Entwicklung von vorausschauenden, selbstoptimierenden Regelverfahren, die gleichzeitig den Nutzerkomfort erhöhen und die Energieeffizienz verbessern.

    \item ID 662\footnote{This use case is no longer available online}: Technologieorientierte Unternehmen stehen nicht nur in der Textilindustrie vor der Herausforderung, aus der Informationsflut systematisch relevantes Wissen für Technologie- und Geschäftsentwicklung herauszufiltern. Mithilfe des funktionssemantischen Radarsystems \enquote{futureTEX} können sie schnell neue Technologien, interessante Anwendungsfelder, Wettbewerber und Experten (Ansprechpartner) identifizieren. Das zentrale Element stellt der mithilfe maschineller Lernverfahren kontinuierlich erweiterte Thesaurus für die Textilbranche dar, der neben technischen Verfahren, Materialien, Eigenschaften, Attributwerten, Institutionen, Orten, Personen auch Kundenanforderungen und Muster zur Erkennung von Anwendungsfeldern enthält. Der Datenbestand umfasst aktuell ca. 200.000 Dokumente aus Forschungsdatenbanken, Patentdatenbanken sowie dem Internet und wird fortlaufend erweitert.

    \item ID 670\footnote{\url{https://www.plattform-lernende-systeme.de/anwendung.html?AID=97}}: Für eine nutzerzentrierte Wissensvermittlung müssen Computersimulationen und digitale Lernspiele den Erfahrungs- und Wissensstand und die intrinsische Motivation der Nutzer berücksichtigen. Eine externe, adaptive Lernkomponente muss die Nutzer im sogenannten Flow-Kanal halten, ausbalanciert zwischen individuellen Herausforderungen und Fähigkeiten. Zur Realisierung werden Verfahren des maschinellen Lernens genutzt, etwa für eine maschinelle Erfassung und Analyse des Lernerzustands, für automatisches Szenenverstehen oder für die automatische Selektion und Erzeugung passender Inhalte.

    \item ID 679\footnote{\url{https://www.plattform-lernende-systeme.de/anwendung.html?AID=88}}: Von einem Angriff auf IT-Systeme können Milliarden von Anwendern, Geräten, Komponenten oder Installationen betroffen sein. Die Realzeitauswertung der schieren Datenmenge potenziell sicherheitsrelevanter Ereignisse braucht Technologien der Künstlichen Intelligenz und des Cognitive Computing. Methoden der Cognitive Security können \enquote{gutartiges} abweichendes Verhalten von \enquote{bösartigem} abweichendem Verhalten im Netz, vor allem im Internet, effektiver und effizienter unterscheiden und den Verteidiger auf die Ausnutzung ihm unbekannter Zero-Days-Schwachstellen hinweisen.

    \item ID 682\footnote{This use case is no longer available online}: Markenpräferenzen und Kaufentscheidungen werden heute mehr denn je durch die eigene Persönlichkeit beeinflusst: Man kauft, womit man sich identifiziert. Fraunhofer-Forscher entwickeln Technologien, die es Unternehmen erlauben, auf die Bedürfnisse des einzelnen Konsumenten einzugehen, ohne seine Privatsphäre zu verletzen. Ein Anwendungsfeld sind digitale Werbetafeln im öffentlichen Raum: Die BioLens-Software des Fraunhofer IGD ermöglicht es, Alter, Geschlecht oder den Grad der Aufmerksamkeit eines Betrachters zu ermitteln und das Werbeangebot in Echtzeit anzupassen. Die Technologien werten Besucherströme und Stimmungslagen statistisch und automatisiert aus. Im System ist hinterlegt, welche Werbung für welche Zielgruppen gezeigt wird.

\end{itemize}

\end{document}